\RequirePackage{fix-cm}
\documentclass[twocolumn]{svjour3}          % twocolumn
\smartqed  % flush right qed marks, e.g. at end of proof
\usepackage{graphicx}
\usepackage{fancyhdr}
\usepackage[english]{babel}
\usepackage[utf8]{inputenc}
\usepackage{graphicx,url,amsmath,subcaption,amsfonts,lipsum}
\usepackage[numbers]{natbib}
\begin{document}

\title{A Complex Network Approach for Nanoparticle Agglomeration Analysis in Nanoscale Images}

\titlerunning{A Complex Network Approach for Nanoparticle Agglomeration Analysis in Nanoscale Images}        % if too long for running head

\author{Bruno Brandoli Machado \and Leonardo Felipe Scabini \and Jonatan Patrick Margarido Oruê  \and Mauro Santos de Arruda \and Diogo Nunes Gonçalves \and Wesley Nunes Gonçalves \and Raphaell Moreira \and Jose F Rodrigues-Jr}

\authorrunning{Bruno Brandoli et. al.} % if too long for running head

\institute{$\bullet$ Bruno Brandoli Machado, Leonardo F Scabini, Jonatan P M Orue, Mauro S Arruda, Diogo N Goncalves and, Wesley N Goncalves  \at
			  CS Dept, Federal University of Mato Grosso do Sul\\
              79907-414, Ponta Pora, MS, Brazil \\
              %Tel.: +556734371743\\
              \email{\{bruno.brandoli,leo.scabini,jonatan.orue,m.arruda,\\diogo.goncalves,wesley.goncalves\}@ufms.br}
           \and
           $\bullet$ Raphaell Moreira \at
              Freie Universitat BerlinTakustr\\
              3, 14195, Berlin, Germany\\
              moreira.raphaell@fu-berlin.de
           \and
           $\bullet$ Jose F Rodrigues-Jr \at
           CS Dept, University of Sao Paulo\\
           13566-590, Sao Carlos, SP, Brazil\\
           junio@usp.br
}

\date{Received: date / Accepted: date}

\maketitle

\begin{abstract}
Complex networks have been widely used in science and technology because of their ability to represent several systems. One of these systems is found in Biochemistry, in which the synthesis of new nanoparticles is a hot topic. However, the interpretation of experimental results in the search of new nanoparticles poses several challenges. This is due to the characteristics of nanoparticles images and due to their multiple intricate properties; one property of recurrent interest is the agglomeration of particles. Addressing this issue, this paper introduces an approach that uses complex networks to detect and describe nanoparticle agglomerates so to foster easier and more insightful analyses. In this approach, each detected particle in an image corresponds to a vertice and the distances between the particles define a criterion for creating edges. Edges are created if the distance is smaller than a radius of interest. Once this network is set, we calculate several discrete measures able to reveal the most outstanding agglomerates in a nanoparticle image. Experimental results using images of Scanning Tunneling Microscopy (STM) of gold nanoparticles demonstrated the effectiveness of the proposed approach over several samples, as reflected by the separability between particles in three usual settings. The results also demonstrated efficacy for both convex and non-convex agglomerates.
\keywords{nanoparticle cluster \and agglomeration analysis \and complex networks}
\end{abstract}

\section{Introduction}

Synthetic nanoparticles have been widely investigated because of their applicability, including drug delivery in medicine \cite{sugaharaCC2009}, cancer treatment and diagnostic tools \cite{kongNANO2011}, and industrial products, such as cosmetics \cite{lorenzN2011}, semiconductor and photovoltaics \cite{eminSE2011}, to name a few. They are designed to have special physical and chemical properties that reflect in their structural characteristics and interaction \cite{masciangioliEST2003}. However, the development and use of new nanoparticles are still constrained by the lack of specialized tools to interpret experimental results, so to characterize such particles \cite{dingSPIE2015}. A particular line of investigation is the safety to human beings. This is because they are more chemically reactive and bioactive, penetrating organs and cells easily. Actually, toxicological studies \cite{liJN2012} have shown that some nanoparticles are harmful to humans.

To better understand the impact of real and synthetic nanoparticles, material scientists use analytical devices whose output are grayscale images of nanoparticles. After the synthesis process and imaging of the nanoparticles, an important task is to extract measurements from such images. Hasselhov and Kaegi \cite{hassellovEHHIN2009} described key visual characteristics that need to be assessed, including \textit{concentration}, \textit{particle size distribution}, \textit{particle shape}, and \textit{agglomeration}. 
Despite the importance of nanoparticle assessment, there is a limited number of works on the characterization of nanoparticles by means of image analysis.

Fisker \textit{et al.} \cite{fiskerJNR2000} developed an automatic method to estimate the \textit{particle size distribution} based on a deformable ellipse model applied to ferromagnetic ($a$-$Fe_{1-x}$-$C_x$) and hematite ($\alpha$-$Fe_{2}O_3$) nanoparticles. In the work of Park \textit{et al.} \cite{parkIIE2012}, the authors propose a semi-automatic method to perform \textit{shape analysis} over the particles. In the work of Park \textit{et al.} \cite{parkIIE2012}, the authors used six images of Transmission Electron Microscopy (TEM) to characterize the shape of gold nanoparticles by representing boundary corners into a parametric curve. Although the authors created a rotation invariant approach, the reconstruction depends on the corners that the algorithm detects. This idea is sensitive to the edges detected by Cany's algorithm \cite{cannyPAMI1986}. Furthermore, since the border detection fails for a number of cases, they reconstruct the particles with incomplete boundaries using functional-PCA (FPCA) \cite{jonesFPCA1992} and the gravity center of each shape. Since the dimensionality is very high, the authors used the curve representation to reduce the number of features with a multidimensional projection method, named Isomap \cite{tenenbaumISOMAP2000}. Finally, they classified the shapes using graph-based clustering and a k-nearest neighbors method over incomplete boundary information. Although this approach presented good accuracy for nanoparticle shape recognition, they did not focus on analyzing the groups and interaction of particles. Vural and Oktay \cite{vuralSIU2014} proposed a method to segment $Fe3O4$ nanoparticles in TEM images by using Hough transform \cite{dudaHOUGH1972}. Similarly, a number of other works \cite{muneesawangSAM2015} used a multi-level image segmentation for measuring the \textit{size distribution} of nanoparticles in TEM images. However, these works disregarded the agglomeration of particles.

We can find a rich literature with similar nanoparticle problems in biomedical imaging, such as detection and counting of cells \cite{liaoNEURO2015}, morphological cell classification \cite{chenCMMM2012}, and cell tracking \cite{zhangBBE2015}. Unlike conventional cell image analysis, agglomeration and interaction analysis of nanoparticles are still a visual counting task. Such task not only demands an extensive work, but it is time-consuming. Therefore, modeling the relationship of nanoparticles in images has emerged as an interesting line of research to characterize their interaction and agglomeration. In this scenario, complex networks define a promising model to draw the relationships observed in nanoparticle images, fostering the comprehension of complex phenomena, most notably interaction, and agglomeration.

\section*{Related Works on Complex Networks}
Complex networks (CN) have emerged as a highly-active research field in the first decade of the XXI century. It came as an intersection between graph theory and probability, resulting in a truly multidisciplinary field, building on top of mathematics, computer science, and physics, leading to a large range of applications. Complex networks are natural structures that represent many real-world systems; its popularity comes from the fact that it is able to model a large range of phenomena. As an illustration, we can cite three developments that have contributed to the research on complex networks: (i) investigation of the random network model \cite{randomCNevolution}; (ii) investigation of small-world networks \cite{smallwordCN}; and (iii) investigation of scale-free networks \cite{scalefreeCN}. Recently, works have focused on the statistical analysis of such networks in order to characterize them.

Complex networks have become an important topic in science due to their ability to model a large number of complex systems such as interaction in society \cite{eustace2015}, processes in biology as protein interaction \cite{barabasiNRG2004}, financial markets \cite{peron2012}, computer vision \cite{goncalvesNEURO2015}, and physics \cite{GoncalvesChaos2012}. In computer science, complex networks have been used to understand the topology and dynamics of the Internet \cite{tylerCT2003}, characterization of social networks \cite{kimTL2015}, text summarization \cite{antiqueiraIS2009}, aspects of scientific co-authorship \cite{newmanCN2004}, and citation networks \cite{alanS2009}.

\section*{Overview of Our Proposal}
Benefiting from the potential of complex networks, we propose a new approach to analyzing nanoparticle agglomeration. As far as we know, this work is the first to report the use of complex networks on nanoparticle images. In the proposed approach, similarly to the work of Fisker {\it et al.} \cite{fiskerJNR2000}, each particle of a nanoscale image is initially detected using 2D-template matching, described in more details in by Brunelli \cite{brunelli2009}. Then, each particle is mapped to a vertex of the complex network. Systematically, a network is built by connecting each pair of nodes by using a threshold for density estimation over a certain radius. For each nanoparticle, we calculate its density, according to which two particles are linked only if the distance between them is lesser than a radius and its density is higher than a given threshold. Then, we represent our complex network topology by calculating the spatial average degree, and the max degree for networks, transformed by different values of radius and thresholds. We tested our approach on real-image particles taken with Scanning Tunneling Microscopy (STM), a technique that creates high-resolution images of nanoparticle settings.

This paper is organized as follows. Section 2 presents a brief review of the complex network theory. In Section 3, the proposed approach for nanoparticle characterization is described in detail. The experiments conducted and the discussions of the results are presented in Section 4. Finally, conclusions are given in Section 5.

\section{Complex Networks}

\subsection{Overview}
In general, works using complex networks have two steps: (i) model the problem as a network; and (ii) extract topological measures to characterize its structure. As complex networks are represented by graphs, every discrete structure such as lists, trees, networks, and images can be suitably modeled. In this context, the main step is to define the best approach to represent the given problem as a set of vertices and connections, so that its complex behavior can be measured as a CN.

\subsection{Complex Networks Representation and Measures}

Complex networks are represented by graphs. An undirected weighted graph $G=\{V, E\}$ is defined wherein $V=\{v_1, ..., v_n\} $ is a set of $n$ vertices and $E=\{e_{v_i,v_j}\}$ is a set of edges connecting two vertices; $e_{v_i,v_j}$ represents the weight of the connection between the vertices $v_i$ and $v_j$.
There are many measures that can be extracted from a CN to characterize it. The reader may refer to the work of Costa \textit{et al.} \cite{cnusp} for a review of different classes of measures. We focused in two simple and important characteristics extracted from each vertex, the degree and the strength. The degree of a vertex $v_i$ is the number of its connections:

\begin{equation}
k(v_{i}) = \sum_{e_{v_i, v_j} \in E} 1
\label{degree}
\end{equation}

The vertex strength is the sum of the weights of its connections:

\begin{equation}
s(v_{i}) = \sum_{e_{v_i, v_j} \in E} e_{v_i, v_j}
\label{eq:STR}
\end{equation}

The vertex degree and strength describe the interaction with neighboring vertices and can be used to analyze the network structure. Globally, it is possible to characterize the behavior of the vertices of the network using the mean degree:

\begin{equation}\label{eq:meanDegree}
\mu_k = \frac{1}{|V|}\sum_{v_{i} \in V} k(v_{i}) 
\end{equation}
and the mean strength:

\begin{equation}\label{eq:meanSTR}
\mu_s = \frac{1}{|V|}\sum_{v_{i} \in V} s(v_{i}) 
\end{equation}

In this work, we analyze the degree and the strength of the vertices to detect regions with strong connections, which are evidence of vertices agglomerates. In the context of our application, these regions present nanoparticle agglomeration, which is the focus of the work.

\section{Proposed Approach for Detection and Agglomeration Analysis}

In this section, we describe our approach for detection and characterization of nanoparticle agglomerates. For this purpose, we use template matching to detect the positions of nanoparticles in nanoscale images. Subsequently, we build a CN with their relative positions. Finally, the degree and strength of the resulting network are used as features to support analysis.

\subsection{Detection of the particles' coordinates}

In order to detect the coordinates of the nanoparticles, we use the template matching technique \cite{brunelli2009}. This technique uses a convolution mask tailored to a feature of interest; this mask corresponds to the {\it template}, which must carry visual characteristics similar to those of what we want to detect. The output of the convolution will be high in the regions of the image whose structure matches the template; the idea is to multiply the image values by large template values -- when there is a match the product gets very high magnitudes compared to the other parts of the image.
The template is constructed by picking a part of a sample image that contains the pattern of interest -- in our case, we picked a well-defined nanoparticle $T(x_t, y_t)$, where $(x_t, y_t)$ represents the pixels in the template. We refer to a given search image as $S(x, y)$. The convolution, then, is performed by moving the center of the template $T(x_t, y_t)$ over each pixel $S(x, y)$ of the image, calculating the sum of products between the coefficients of $S$ and $T$ over the whole area spanned by the template. After the convolution, the positions with the highest scores will correspond to the patterns of interest -- in our case, the set of nanoparticles.

\subsection{Modeling Complex Networks for Nanoparticle Agglomeration Analysis}

The CN is built after the spatial positions of the nanoparticles in the image. The network is built considering each nanoparticle as a vertex. To build the set $E$, the weight of the connections is defined according to the Euclidean distance -- shortly referred to as a function $dist:V\times V \rightarrow \mathbb{R}$. In order to connect only close vertices, a radius $r \in [0, 1]$ is considered. First, the edges' weights, $e_{v_{i}, v_{j}}$, are normalized into the interval $[0, 1]$ dividing its Euclidean distance $dist_{v_{i}, v_{j}}$ by the distance between the two more distant vertices, as follows:

\begin{equation}
e_{v_{i}, v_{j}} = \frac{\sqrt{(x_{i}-x_{j})^{2}+(y_{i}-y_{j})^{2}}}{max(dist_{v_{i}, v_{j}})}
\end{equation}
where $x_i$ and $y_i$ are the spatial coordinates of the nanoparticles and $max(dist_{v_{i}, v_{j}})$ is the distance between the two most distant nanoparticles.

Then, the connection between each pair of vertices is maintained if its normalized Euclidean distance $e_{v_{i}, v_{j}}$ is less or equal to $r$. Moreover, we complement the normalized weight $e_{v_{i}, v_{j}}$ with respect to the threshold radius $r$, as follows:

\begin{equation}
e_{v_{i}, v_{j}}=\left\{
\begin{array}{l}
r - e_{v_{i}, v_{j}}$, \ \ \ if $ e_{v_{i}, v_{j}} \leq r
\\
\ \ \ 0 $,  \ \ \ \ \ otherwise$
\end{array}
\right
.
\label{eq:complement}
\end{equation}

It is important to notice that $r - e_{v_{i}, v_{j}}$ inverts the edge weight, which was originally the Euclidean distance. After this operation, the closer any two vertices are, the higher is their weight. This is performed considering the vertex strength, that is, stronger vertices represent higher interplays among neighbors.

The resulting CN contains connections between vertices inside a given radius, according to the Euclidean distances. However, this representation does not consider the agglomeration level of the vertices, which is the main purpose of the problem, i.e., the use of a radius to connect close vertices is not sufficient to properly represent agglomerates. To finally model the network in a proper way, revealing the level of nanoparticle agglomeration, we propose another transformation on its topology. A new function is applied to calculate the density of the vertices, which represents their relation to their neighbors in terms of distance. This measure can be calculated using the CN information obtained so far. It becomes necessary to extract the degree $k(v_{i})$ and the strength $s(v_{i})$ of the vertices (Equations \ref{degree} and \ref{eq:STR}), both measures depending on the neighborhood of each vertice. The neighborhood is defined by the radius $r$, so each vertex inside the distance defined by the radius value is analyzed. Given a resulting CN $G^r$ built with a radius $r$, and the respective degree and strength of each vertex $v_i$, its density is defined by:

\begin{equation}
d(v_i) = \frac{s(v_i)}{k(v_i)}
\end{equation}

After calculating the density of each vertice, we normalize the densities so to have a domain of values inside the range $[0 ,1]$.

Since the degree is the number of connections and the strength is the sum of its weights, the density $d(v_i)$ refers to the average weight of a vertice's neighborhood. Following the complement operation defined in Equation \ref{eq:complement}, then, vertices with a larger number of close neighbors tend to have greater densities.

With the density, it is possible to perform another transformation to highlight the agglomerates of the network. We proceed by considering only the connections between vertices with density higher than a threshold $t$, discarding the other ones. In this context, a new CN $G^{r, t}$ is obtained by analyzing each edge $e_{v_{i}, v_{j}}$, as follows:

\begin{equation}
e_{v_{i}, v_{j}}=\left\{
\begin{array}{l}
e_{v_{i}, v_{j}}$, \ \ \ if $d(v_i)$ and $ d(v_j) \geq t
\\
\ \ 0 $,  \ \ \ \ \ \ \ \ \ \ \ \ \ otherwise$
\end{array}
\right
.
\end{equation}

\begin{figure*}[!htb]
	\centering
	\makebox[\linewidth][c]{
		\begin{subfigure}{0.45\linewidth}
			\centering
			\includegraphics[width=0.75\linewidth]{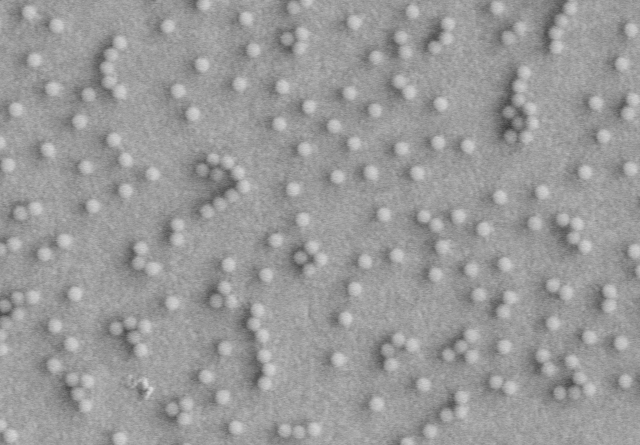}
			\caption{Input image.}
		\end{subfigure}	
		\begin{subfigure}{0.45\linewidth}
			\centering   	
			\includegraphics[width=0.95\linewidth]{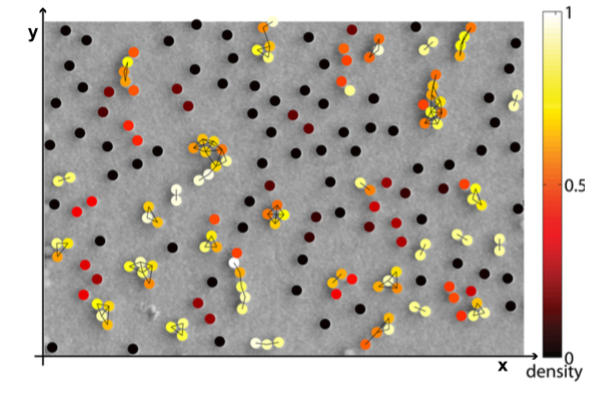}    	
			\caption{Complex Network built using the nanoparticle positions (absolute coordinates).}
		\end{subfigure}
	} 
	\caption{Nanoparticle image modeled as a Complex Network according to the proposed approach. (a) Input image. (b) Density of each nanoparticle (color-mapped) and connections of the resulting Complex Network.}
	\label{fig:sample1}
\end{figure*}

This final transformation results in a CN that better represents the agglomeration of the vertices, instead of the limited distance analysis of the first transformation. It means that the use of the density to define connections allows selecting edges in regions of interest, i.e., with high density.
In the context of the current application, the network now presents connections between nanoparticles that pertain to agglomerates; these connections come according to the radius $r$ and to the density threshold $t$. %Notice that with the current computational technologies, both $r$ and $t$ allows for an interactive analysis; with the user iteratively defining agglomerates according to his/her needs.
In Figure \ref{fig:sample1}, the positions in a real image of nanoparticles is analyzed and a CN is modeled using radius $r=0.04$ and threshold $t=0.5$. The color indicates the density ranging from black/red (low density) to white/yellow (high density).

%\begin{figure*}[!htb]
%	\centering	
%	\subfigure[Input image]{\includegraphics[width=0.75\linewidth]{original}}
%	\\
%	\subfigure[Complex Network built using the nanoparticle positions (absolute coordinates).]{\includegraphics[width=0.95\linewidth]{eixos.png}}
%	
%	\caption{\label{fig:sample1} Nanoparticle image modeled as a Complex Network according to the proposed approach. (a) Input image. (b) Density of each nanoparticle (color-mapped) and connections of the resulting Complex Network.}
%\end{figure*} 

\subsection{Dynamic Analysis of Complex Networks}
\label{subsec:analysis}
\begin{figure*}[!htb]
	\centering
	\makebox[\linewidth][c]{
		\begin{subfigure}{0.45\linewidth}
			\centering
			\includegraphics[width=\linewidth]{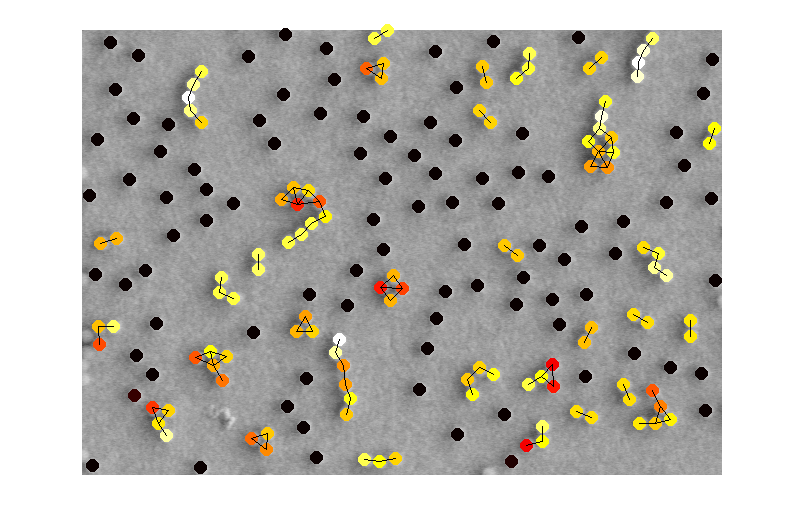}
			\caption{$r=0.03$, $t=0.2$.}
		\end{subfigure}
		
		\begin{subfigure}{0.45\linewidth}
			\centering   	
			\includegraphics[width=\linewidth]{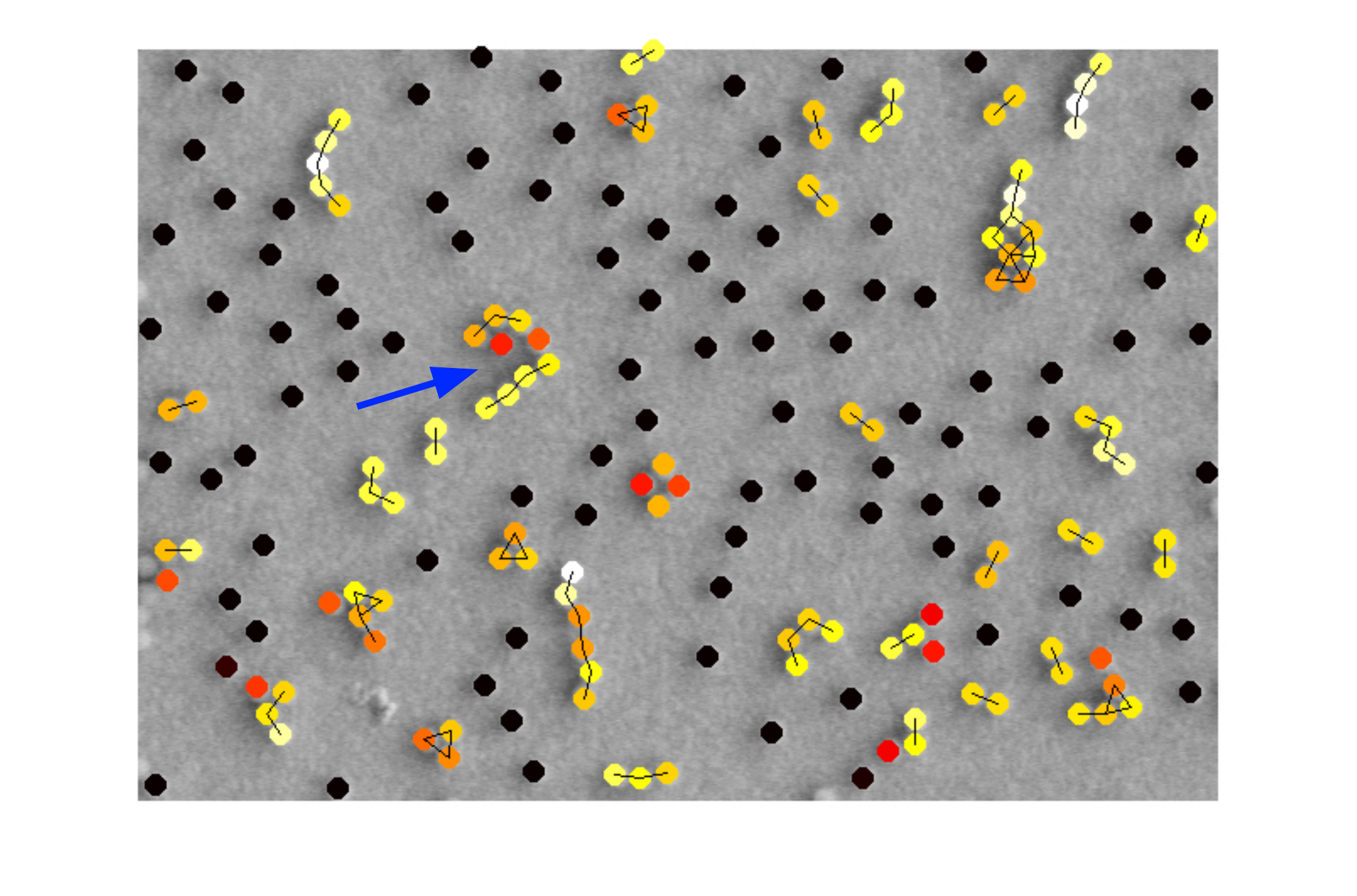}  	
			\caption{$r=0.03$, $t=0.5$.}
		\end{subfigure}
	}
	\makebox[\linewidth][c]{
		\begin{subfigure}{0.45\linewidth}
			\centering
			\includegraphics[width=\linewidth]{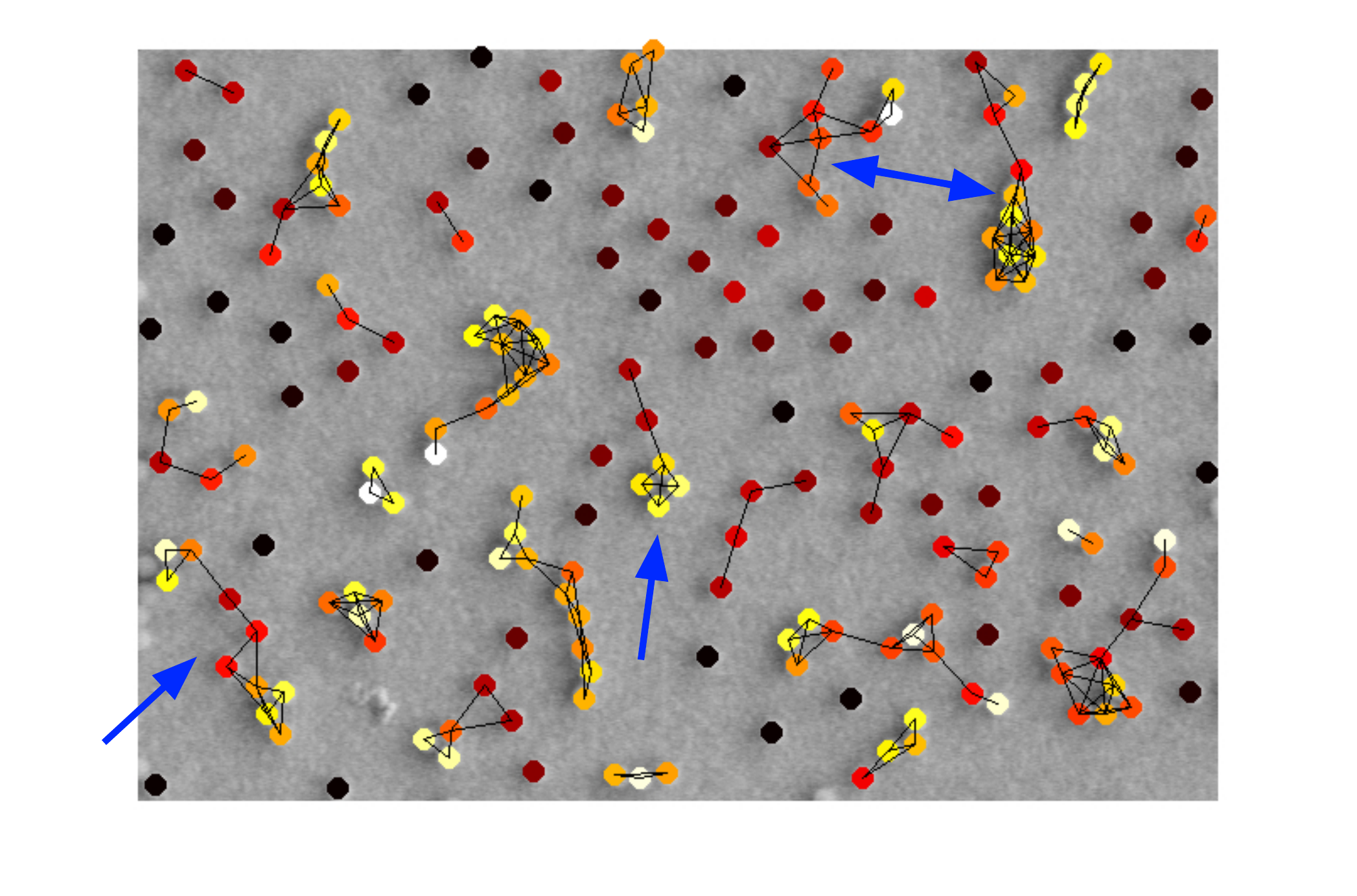}
			\caption{$r=0.05$, $t=0.2$.}
		\end{subfigure}
		
		\begin{subfigure}{0.45\linewidth}
			\centering   	
			\includegraphics[width=\linewidth]{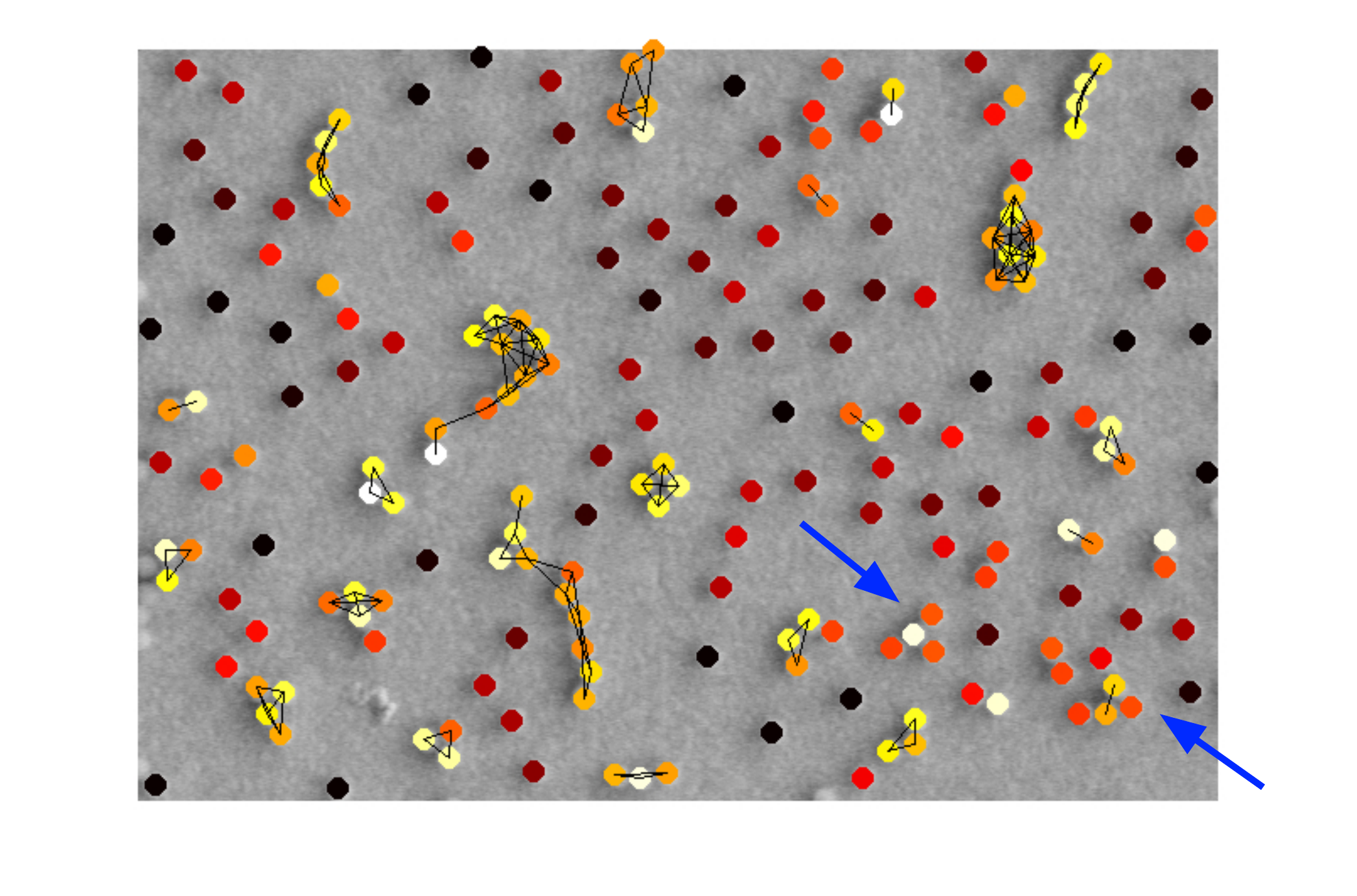}  	
			\caption{$r=0.05$, $t=0.5$.}
		\end{subfigure}
	} 	
	\caption{Complex Network topology changes by varying the parameters $r$ and $t$. Blue arrows correspond to the most relevant areas affected by changing the parameter values. Basically, if $r$ and $t$ are increased, new edges are created or removed, depending on the intrinsic agglomeration.}
	\label{fig:dynamics}
\end{figure*}

To analyze a nanoparticle image, it is necessary to consider its corresponding CN in view of the range of parameters that influence the formation of agglomerates. As mentioned earlier, we consider two parameters, the radius $r$ and the threshold $t$, which affect the network topology (set of edges) resulting in networks with dense or sparse connections -- illustrated in Figure \ref{fig:dynamics}. This variable configuration is useful to analyze the network considering different analytical demands -- it is possible, for instance, to consider bigger or smaller agglomerates, denser or sparser, separated or closer, depending on the material and on the problem at hand. In fact, a network characterization cannot be fully complete without considering the interplay between structural and dynamic aspects \cite{structureanddynamics}.

Therefore, to perform a thorough analysis, one must take into account a set of radii $R=\{r_1, ..., r_{nr}\}$ and a set of thresholds $T=\{t_1, ..., t_{nt}\}$ able to characterize the network in the amplitude of parameters $r$ and $t$. To do so, we build multiple topologies, each one given by a combination of $r$ and $t$. The problem, then, becomes how to put these multiple topologies together in a coherent mathematical representation. We do it my means of a feature vector whose dimensions are given by measures extracted from each $r-t$ Complex Network.

% The resulting networks are evaluated individually, i.e., each topology provides different measures that, combined, result in a robust feature vector. Thus, the network growth is evaluated from its creation (little values) until its stabilization (high values). This kind of analysis allows a complex characterization of the nanomaterial, considering the global structure of the nanoparticles.

\subsection*{Feature Vector}
Given two sets, $R=\{r_1, ..., r_{nr}\}$ and $T=\{t_1, ..., t_{nt}\}$, we build $|R| \times |T|$ CNs, each one denoted $G^{r,t}$. From each CN we calculate four measures: mean degree (Equation \ref{eq:meanDegree}), max degree $k_{max} = \{k(v_i)|k(v_i)>k(v_j),i \neq j,\ \forall\ v_i \in V, v_j \in V\}$, mean strength (Equation \ref{eq:meanSTR}), and max strength $s_{max} = \{s(v_i)|s(v_i)>s(v_j),i \neq j,\ \forall\ v_i \in V, v_j \in V\}$. 
Finally, the feature vector, denoted $\varphi$, is formed by the concatenation of the sequence of  four measures of each CN $G^{r,t}$, as follows:

\begin{equation}
\varphi = [\mu _k ^{r_1, t_1}, \ k_{max}^{r_1, t_1}, \ ..., \ \mu k ^{r_{nr}, t_{nt}}, k_{max}^{r_{nr}, t_{nt}}]
\end{equation}

The number of features depends on the number of radii, thresholds, and extracted measures; its cardinality is given by $|\varphi| = nr*nt*m$, where $nr$ is the number of radii, $nt$ is the number of threshold values, and $m$ is the number of measures. For a consistent domain of values considering any given vectors, the features are numerically homogenized according to the standard score~\cite{larsen2012} technique; that is, from each feature we subtract the mean score and divide the result by the standard deviation of all features.

\section{Results and Discussion}
\label{sec:experiments}

In this section, we evaluate the proposed approach. We show results in real nanoparticle images by differentiating three different cases of agglomeration.

\subsection{Image Dataset}
\label{sec:dataset}

\begin{figure*}
	\centering
	\makebox[\linewidth][c]{
		\begin{subfigure}{0.3\textwidth}
			\centering
			\includegraphics[width=0.9\textwidth]{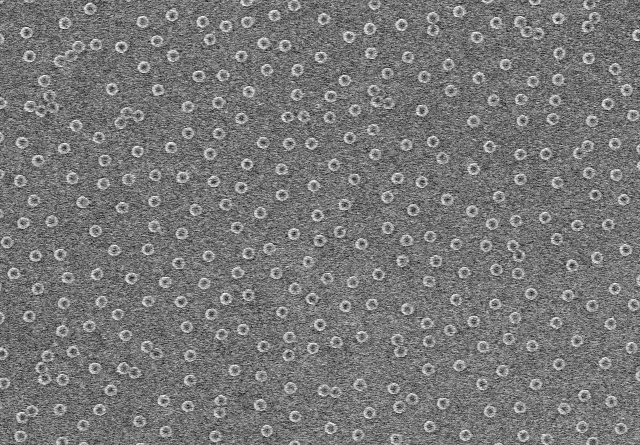}
			\caption{Case 1.}
		\end{subfigure}
		
		\begin{subfigure}{0.3\textwidth}
			\centering
			\includegraphics[width=0.9\textwidth]{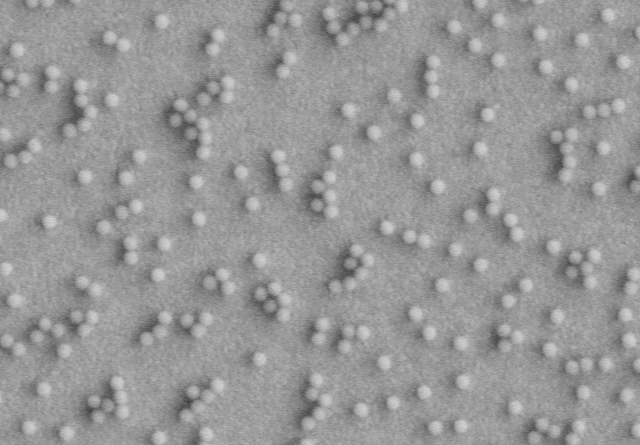}
			\caption{Case 2.}
		\end{subfigure}
		
		\begin{subfigure}{0.3\textwidth}
			\centering
			\includegraphics[width=0.9\textwidth]{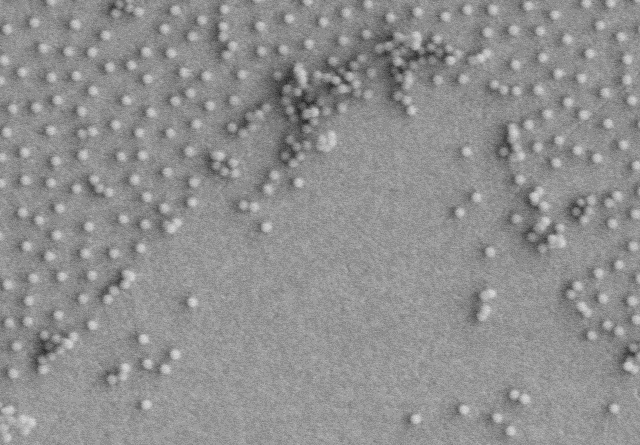}
			\caption{Case 3.}
		\end{subfigure}		
	}
	\caption{Images for the three levels of nanoparticle agglomeration used in the experiments.}
	\label{fig:images}	
\end{figure*}

We built a dataset of nanoparticle images with 10 samples labeled into 3 agglomeration cases: Case 1 -- the images have few groups and the nanoparticles are uniformly spread; Case 2 -- the number of groups is larger if compared to Case 1, with little nanoparticle agglomeration; Case 3 -- the images have a strong level of agglomeration and overlapping. Each kind of image can be observed in Figure \ref{fig:images}. Notice that nanoparticle agglomeration might happen in the 3D scenario, with a strong incidence of overlapping; since our method works over 2D images, such cases are to be tackled with alternative techniques (e.g., 3D reconstruction \cite{van2011determination}), alternatively, the samples shall be prepared according to a laboratory protocol that reduces overlapping. For our experiments, we have used STM images of gold nanoparticles; standard reference materials NIST 8011, 8012, and 8013 – NIST$^\circledR$, Gaithersburg, MD, U.S. The gold particles were suspended in a solution of deionized (DI) water at a concentration of 250,000 particles/mL. In order to avoid dissolution of the gold nanoparticles, acid was not added. 

\subsection{Assessing Parameters}

A problem that raises in our methodology refers to the choice of the best parameters $r$ and $t$. These two parameters render a set of possibilities large enough to impede the user find the best configuration. We treat this issue by measuring the quality of the agglomerations detected by each pair $(r,t)$; to do so, we use the well-known measure named \emph{silhouette coefficient}~\cite{kumar2005}, which was originally proposed to evaluate clustering algorithms.

In our setting, the silhouette coefficient shall measure the cohesion and the separation between the agglomerates detected in a given $G^{r,t}$ configuration. Considering a nanoparticle $v_i$ belonging to an agglomerate, its cohesion $a_{v_i}$ is given by the average of the distances between $v_i$ and all the other nanoparticles belonging to the same agglomerate. In turn, the separation $b_{v_i}$ is given by the smallest distance between $v_i$ and all the other nanoparticles belonging to the other agglomerates. Once we calculate the cohesion and the separation of a given partice $v_i$, its silhouette value is given by:

\begin{equation}
silhouette(v_i) = \frac{b_{v_i} - a_{v_i}}{\max(a_{v_i},b_{v_i})}
\label{eq:silhouette}
\end{equation}

Given a set with $n$ instances and a corresponding clustering, the silhouette of the whole set is given by the average of the silhouette (Equation \ref{eq:silhouette}) of all of its instances. The average silhouette, Equation \ref{eq:avgsilhouette}, provides a number that characterizes how good is the clustering (set of agglomerates).

\begin{equation}
S = \frac{1}{|V|}\sum_{v_{i} \in V} silhouette(v_i)
\label{eq:avgsilhouette}
\end{equation}

The silhouette can range between $-1 \leq S \leq 1$, where larger values indicate better cohesion and separation between agglomerates, that is, better agglomeration. Negative values indicate instances (nanoparticles) assigned to the wrong agglomerate; this is because the distances indicate that particles from other agglomerates are closer than the particles of the agglomerate to which they were assigned. 
%In our experiments, clusters are composed taking into account labeled instances, and the silhouette indicates whether images belonging to the same class are more similar between themselves than images belonging to other classes. Therefore, the best set of features is the one which yields the projection with the largest silhouette coefficient.

\subsection{Definition of the best values for $r$ and $t$}
\label{subsec:radius}
\begin{figure}[!htb]
	\centering
	\includegraphics[width=0.95\linewidth]{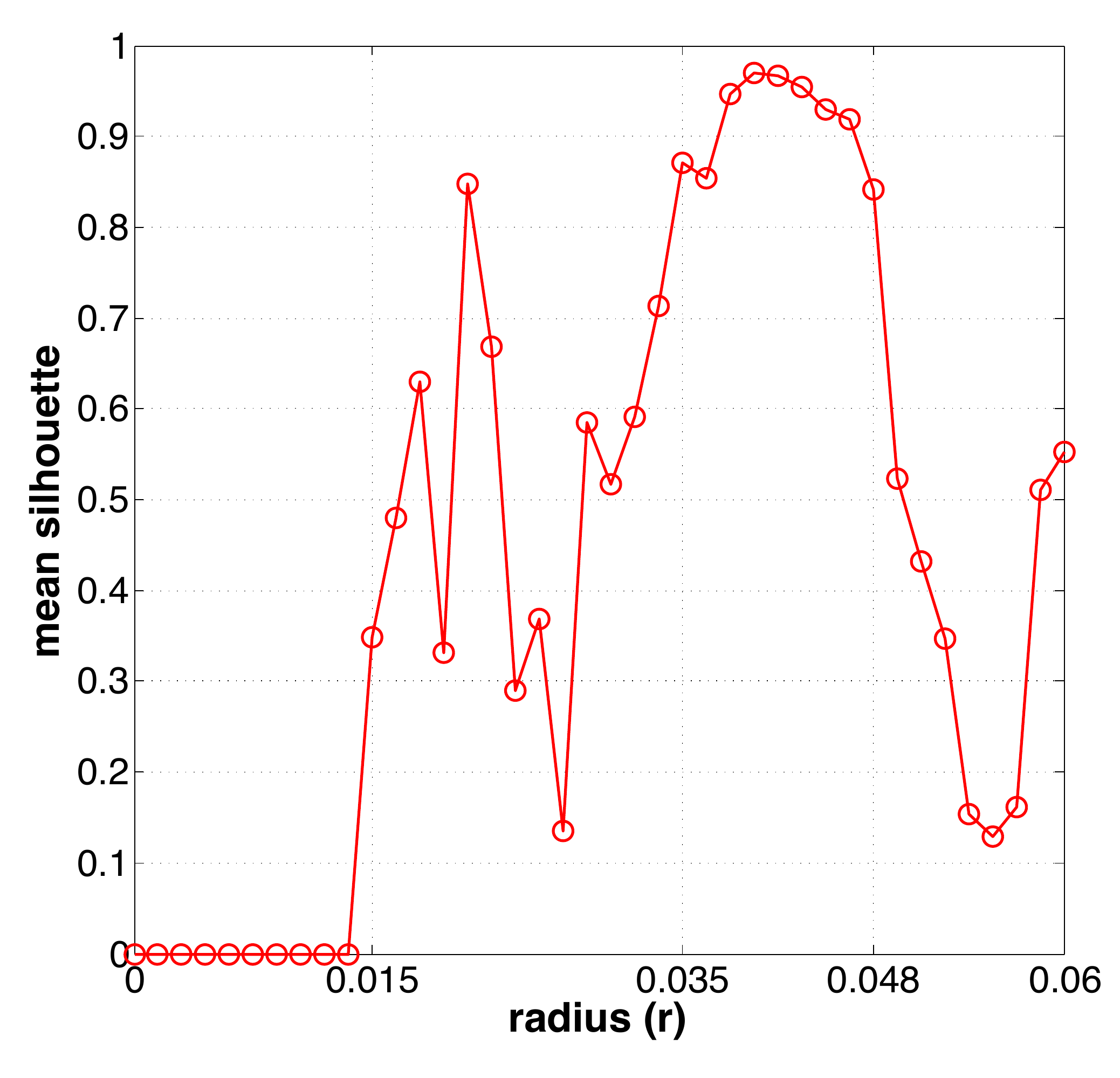}
	\caption{\label{fig:rinterval} Silhouette value in function of the radius $r$.}
\end{figure}

We defined the best interval of values by analyzing the average silhouette (Equation \ref{eq:avgsilhouette}) from CNs built with different values for parameters $r$ ant $t$. The radius was analyzed by varying its value in the range $[r_1=0, r_{nr}=0.06]$,
while $t$ was analyzed in the range $[t_1=0.1, t_{nt}=0.9]$. For each combination of values, we calculated the silhouette value for each image. We summarize the multiple silhouette values using mean; hence, the y-axes of our images are labeled {\it mean silhouette}. With this configuration, we note that the domain given by parameters $r$ and $t$ renders a 3D plot, that is, a surface of silhouette values. We do not plot the surface, since static 3D images are of little use; instead, we present the best section of the surface in Figure \ref{fig:rinterval}. This section includes the highest silhouette values of the r-t domain.
%XXXX AINDA PRECISAVA DIZER QUAL O VALOR DE t QUE CORRESPONDE À MELHOR SECCÇAO DA SUPERFICIE

The results allow to detect and discard values from the radius interval that do not produce good detection of agglomerates. We can observe that radius values in the range $0 < r < 0.015$ are not sufficient to connect nanoparticles/vertices in the CN. In this settings, it was not possible to identify proper sets of agglomerates; therefore, it was not possible to calculate the silhouette, which explains the silhouette $0$. We also discard values in the range $[0.015, 0.035]$ due to the inconstant results that are observed; although there is a peak at radius $0.0215$. By analyzing other values, it is possible to notice that the best results are achieved in the range $[0.035, 0.048]$, a stable interval of silhouette values. Following, there is a decrease of performance probably caused by CNs with dense connections that are not useful to discriminate the agglomerates. Based on these results, we opt for the radius interval $[r_1=0.03, r_{nr}=0.05]$; accordingly, the radius set $R=\{r_1, ..., r_{nr}\}$ will be composed by $nr$ equidistant values ranging from $0.03$ to $0.05$. Notice that the curve could be smoother had we used more images; this is because with a larger number of values (one per image), the expected average value would dominate the plot, avoiding spikes. However, that would demand a very large number of images, which were not available and which is not usual in the corresponding literature \cite{fiskerJNR2000,parkIIE2012,parkPAMI2013,muneesawangJN2015}.

We defined the threshold best interval of values by analyzing the average silhouette in the range $[t_1=0.1, t_{nt}=0.9]$ using $nt$ equidistant values.

\subsection{Evaluation of Complex Network Measures}

To characterize the CN structure, 4 measures were evaluated: mean degree, max degree, mean strength, and max strength -- refer to Section \ref{subsec:analysis}. We consider each possible combination among these measures to find the one that best reflects the silhouette of the agglomerates identified using the best values of $r$ and $t$, as explained in Section \ref{subsec:radius}. Each result can be observed in Table \ref{table:silhueta} along with the standard deviation of the silhouette value.

\begin{table}[!htb]
	\caption{Mean silhouette in perspective of complex network measures mean degree ($\mu_k$), max degree ($\mu_s$), mean strength ($\mu_s$), and max strength ($s_{max}$). We used parameter values as pointed out by the results of Section \ref{subsec:radius}.}
	\centering
	\begin{tabular}{|c|c|c|c|c|c|}
		\hline
		$\mu_k$ & $k_{max}$ &  $\mu_s$ & $s_{max}$& no. of features & silhouette  \\
		\hline
		X &  &  & & 18 & $0.88$ ($\pm 0.11$)  \\
		\hline
		& X &  & & 18 & $0.15$ ($\pm 0.13$)  \\
		\hline
		& & X & & 18 & \boldmath{$0.89$} \boldmath{($\pm 0.07$)}  \\
		\hline
		& &  & X & 18 & $0.24$ ($\pm 0.35$)  \\
		\hline
		\hline
		X & X &  &  & 36 & $0.39$ ($\pm 0.07$)  \\
		\hline
		X &  & X &  & 36 & \boldmath{$0.91$} \boldmath{($\pm 0.07$) } \\
		\hline
		X &  &  & X & 36 & $0.91$ ($\pm 0.08$)  \\
		\hline
		& X & X &  & 36 & $0.27$ ($\pm 0.11$)  \\
		\hline
		& X &  & X & 36 & $0.27$ ($\pm 0.10$)  \\ 
		\hline
		&  & X & X & 36 & $0.33$ ($\pm 0.18$)  \\ 
		\hline
		\hline
		X& X & X & & 54 & $0.41$ ($\pm 0.07$)  \\ 
		\hline
		X& X &  & X & 54 & $0.40$ ($\pm 0.07$)  \\ 
		\hline
		X&  & X & X & 54 & \boldmath{$0.92$} \boldmath{($\pm 0.07$)}  \\ 
		\hline
		& X & X & X & 54 & $0.28$ ($\pm 0.12$)  \\ 
		\hline
		\hline
		X& X & X & X & 72 & $0.41$ ($\pm 0.08$)  \\ 
		\hline
	\end{tabular}
	\label{table:silhueta}
\end{table}

According to these results, one notices that the max degree and max strength do not provide good discrimination if used individually. However, the max strength proved to be useful if combined with the mean strength and mean degree, producing the best result ($0.92$) using 54 features. The means, individually and combined, provided results close to the best ($0.88$, $0.89$ and $0.91$ combined), but on the other hand, they use fewer features (18 individually and 36 combined). The combination of all the features proved to be not applicable for agglomeration analysis. Finally, it is possible to conclude that measures mean strength ($\mu_s$) and max strength ($s_{max}$), together, have the highest representativeness with respect to the intrinsic agglomerative properties of complex networks derived from nanoparticle images.

\section{Conclusion}
\label{sec:conclusao}

The analysis of nanoparticles agglomeration is a topic of recurrent relevance for the interpretation of experiments in the field of nanomaterials. Hence, in this work, we proposed a novel approach for nanoparticle agglomeration analysis based on complex networks. Our method innovates in the sense that its parameters allow for analyses modeled by the interests of the user, including the material and the problem at hand; besides, it adheres to a visual analysis. During our experiments, we showed how to identify the best configuration of parameters by using the metric of silhouette, usually used in clustering problems. We conducted experiments on three levels of agglomeration so to cover usual settings of experimental environments. The results were quantitatively convincing, demonstrating the feasibility of the method, which can handle a large number of particles at the same time that it is much faster and less subjective than commonly-used manual techniques.

The results support the idea that our approach can be used in nanoparticle analysis in material engineering, improving visual analyses for important industries, such as cancer treatment, cosmetics, pharmaceutics, photovoltaics, and food.

\begin{acknowledgements}
The authors are thankful to the AG workgroup for the STM images. Bruno Brandoli was partially supported by FAPESP under grant 02918-0. The authors also acknowledge the support from FUNDECT.
\end{acknowledgements}

\bibliographystyle{spbasic}      % basic style, author-year citations

\end{document}